\documentclass[final,5p,times,twocolumn]{elsarticle}
\usepackage[utf8]{inputenc}
\usepackage{epsfig}
\usepackage{amssymb}
\usepackage{amsmath}
\usepackage{braket}
\usepackage{enumerate}
\usepackage{hyperref}
\usepackage{cleveref}
\journal{Physics Letters B}

\begin{document}
\begin{frontmatter}
\title{Broken axial symmetry as essential feature for a consistent~modelling 
of~various~observables in heavy nuclei}

\author{E. Grosse}

\ead{e.grosse@tu-dresden.de}
\address{Institute of Nuclear and Particle Physics, Technische Universit\"at Dresden,
01062 Dresden, Germany} 

\author{A.R. Junghans}
\ead{a.junghans@hzdr.de}

\address{ Institute of Radiation Physics,  Helmholtz-Zentrum Dresden-Rossendorf,  01328 Dresden, Germany }


\begin{abstract}

Although most nuclear spectroscopy as well as atomic hyperfine structure data do not deliver accurate information on nuclear axiality the ad-hoc assumption of symmetry about one axis found widespread use in nuclear model calculations. In the theoretical interpretation of nuclear properties as well as in the analysis of experimental data triaxiality was considered  -- if at all --  only for some, often exotic, nuclides. A breaking of axial symmetry combined to a spin-independent moment of inertia results in a surprisingly simple heuristic triaxial parametrization of the yrast sequence in all heavy nuclei, including well deformed ones. No additional fit parameters are needed in detailed studies of the mass and charge dependence of the electric dipole strength in the range of and outside of giant dipole resonances. Allowing triaxiality also avoids the introduction of an arbitrary level density parameter ã to fit the accurate values observed in n-capture experiments and ã can be taken from nuclear matter studies. A combination of this value to the yrast energies no longer based on axiality and the related I(I+1) rule results in agreement to data independent of spin. And predictions for radiative neutron capture as derived on the basis of non-axiality are improved as well. The experimentally favoured broken axial symmetry is in accord to HFB and MC-shell model calculations already for nuclei in the valley of stability.
\end{abstract}
\end{frontmatter}
\section{Introduction}
\label{sec:1}

About 80 years ago evidence was presented \cite{ja37} for an energy reduction in configurations of reduced symmetry and this Jahn-Teller effect is important for the physics of large molecules. At about the same time the hyper-fine structure observed in atomic spectra \cite{sc35} suggested the concept of nuclear deformation, i.e. the breaking of spherical symmetry, but no atomic information on broken axial symmetry in nuclei was available. Thus an apparent accord of about 30 ground state band energies in rather well deformed nuclei to the I(I+1)-rule lead to the assumption \cite{bo53, mo57} of symmetry about one axis for them. Many studies were performed \cite{er58, da64, bo75} to relate nuclear observables to the concept of axiality. Most analysis of experimental findings assumed the nuclear cores to be either spherical or axially symmetric. Finite Range Liquid Drop model (FRLDM) calculations based on a microscopic-macroscopic approach for the ground state energies of heavy nuclei have concluded that deviations from axial symmetry have a very small effect in the valley of stability such that nuclei from there can be considered axial  \cite{mo06} in view of calculational uncertainties, and they are not contained in 
respective tables \cite{mo08}. Similarly, an earlier Thomas-Fermi study \cite{du08} of randomly selected nuclei allowing axial symmetry breaking indicates a mass decrease, albeit small, with a non-zero triaxiality parameter $\gamma > 0^\circ $ \cite{bo75, hi53}  in all of them. Apparently ground state masses are only very weakly sensitive to the breaking of axial symmetry. 

Jahn-Teller-considerations can also be applied to axial symmetry and its eventual breaking \cite{re84}. Over decades some respective studies were made by theory or data analysis \cite{df58, da65, dr59}. Early works on odd nuclei \cite{me74, to77} and on the signature change in odd-odd nuclei \cite{be84, ha90} have dealt with broken axiality. Electromagnetic transitions between non-yrast levels \cite{cl86, ma90, wu96, sr06}, regarding the splitting of magnetic strength in heavy nuclei \cite{pa85} and very recent spectroscopic studies of fission fragments \cite{na17, do17} indicate broken axial symmetry. A recent Hartree-Fock-Bogolyubov (HFB) calculation \cite{de10} combined to the generator coordinate method (GCM) predicts a broken axiality for the ground states in at least 1700 of the nuclei studied; in these the large uncertainty of the results does not include zero for the angle $\gamma$. Newer theoretical considerations discuss triaxiality in nuclear rotation by reference to a few nuclides only \cite{al17,sc19,ot22} and a very recent auxiliary-field quantum Monte Carlo (AFMC) simulation \cite{gi18} confirms axial asymmetry for the 9 nuclei studied. 

The work presented here is motivated by the search for observations indicating broken axial symmetry for as many nuclei as investigated to play a role for it. To find indicators apart from the ground state mass \cite{du08} we have reanalysed various recently published data under this aspect. Three topics will be covered: (1) nuclear spectroscopy, (2) level densities, (3) splitting of giant dipole resonances (GDR); these are well known to be sensitive to shape symmetry. We cover the mass number range from 70 to 250, mainly from the valley of stability. In addition to Mo-isotopes, for which we have demonstrated \cite{er10} the effect of non-axiality on GDR shapes already 12 years ago, data from more than 150 heavy nuclei with $50<A<250$ have been reviewed \cite{gr16} and for some of them measurements are reanalysed here. 
We avoid any local parameter-fit to the data in question and our attempt is based on global parameters eventually defined by other observations.

\section{Nuclear Excitation Energies, Data and Models}
\label{sec:2}
The aim of this section is to describe physical trends rather than to present detailed studies for specific nuclei. For decades excited states in odd nuclei \cite{me75} triaxiality was discussed either with respect to the spherical shell model \cite{go55} or reference was made to the seminal work on a “Classification of the Nucleonic States in Deformed Nuclei” \cite{ni55}. In that paper a connection line between states in the two regimes is depicted although the calculations were explicitly \cite{mo55} performed for well deformed nuclei only. Exceptionally, e.g. \cite{la72, es97}, similar calculations were also mentioned for the interpretation of data in the case of smaller quadrupole strength in even nuclei. Here one standard observable is the transition rate $B(E2)$ from ground to the $1^{st}$ excited $2^+$ level. It was proposed \cite{ra01} to derive from it a “transition quadrupole moment" $Q_t$ as a measure of deformation. This avoids the use of parameters as defined for it in various theoretical works -- unfortunately not always in the same way.

For some nuclei, $Q_t$ does not differ much from the quadrupole moment $Q_0$ as measured by Coulomb excitation with reorientation \cite{st05}. These are rather tedious experiments with large uncertainty and probably this is the reason for the habit to identify nuclear quadrupolar shape by $Q_t$. The difference between $Q_0$ and $Q_t$ becomes larger for nuclei with small $Q_t$, in accord to theoretical work \cite{df58, ri82, al17, sc19} allowing broken axial symmetry. A recent study \cite{zh99} on 2$^+$~-levels in medium mass nuclei (Z=50 -- 82) finds an accord between the models of interacting bosons and of triaxial rigid rotation \cite{df58}. For a wider range in $A$ the latter model was applied \cite{ei87} to the excitation energy ratios of the two lowest 2$^+$ states and extracted values for the triaxiality parameter $\gamma$ to be between $7^{\circ}$ and $12^{\circ}$ for several well deformed nuclei – indicating broken axial symmetry also in well deformed nuclei.  

\

In the subsequent discussion of excited states we relate such symmetry breaking to two old observations for a large number of nuclei and confirmed more recently:  

(1) The ratio $R_{42}=E_x(4^+)/E_x(2^+)$ is strongly correlated \cite{gr62, ra88} to $Q_t$ and hence characterises a reduction of collective E2-strength in heavy nuclei when leaving the region of large $Q_t$ , i.e. large deformation. 

(2) The yrast energies in the "ground band" of nearly all heavy nuclei were observed to deviate from the $I(I+1)$-rule \cite{bo75}. This rule was shown for any rotors to always be a direct consequence of axiality \cite{ma56}. If a deviation from it is observed the axial symmetry of such a rotating body is broken or it is not rigid. As centrifugal stretching was mentioned to play a role here we will discuss it later. 

Before that we analyse various experimental facts without assuming axial symmetry. We start by noting that the quantum mechanical treatment of rigid rotation leads to an analytic expression for axial rotors only \cite{ll48}. For non-axial ones this $I(I+1)$-rule has to be replaced by more involved calculations \cite{df58, dr59}. For heavy nuclei we propose to allow for axial symmetry breaking by an approximation to these, which we tested to be good for our parameter range: In a power series in $I$ higher orders are neglected and the quadratic term is modified by an axiality factor $c_x$. And Eq.~\ref{eq:1} explicitly avoids $I(I+1)$:
\begin{equation}
   E_{yr}(I) = B \cdot (I + |c_x| \cdot I^2) ;\  \  \  \
    c_x =\frac{2-R_{42}}{2 \cdot R_{42}-8} + 0.012
\label{eq:1}
\end{equation}
This expression approaches the conventional form $E_{yr}\propto(I+I^2)$ for $c_x\rightarrow 1$ and in the following we use $c_x$ as indicator of reduced axiality and its possible breaking. To test it we make use of recent data compilations \cite{pr17,ra01} which exist for the experimental quantities $E(2+)$, $E(4+)$, and B(E2,$ 0^+ \rightarrow 2^+)$ for more than 300 nuclides. We use 150 of them with A $>$ 50 in this section as well as in the later ones. We derive $c_x$ from $R_{42}$ values for the lowest 2+ and 4+ levels to reexamine the correlation between excitation energies and transition quadrupole moments \cite{gr62,pr17}. To properly approach $Q_t\rightarrow 0$ we add a very small term to $c_x$, adjusted to $E_x (2^+)$ in $^{208}$Pb. 

\begin{figure}[ht]
\includegraphics[width=1.5\columnwidth]{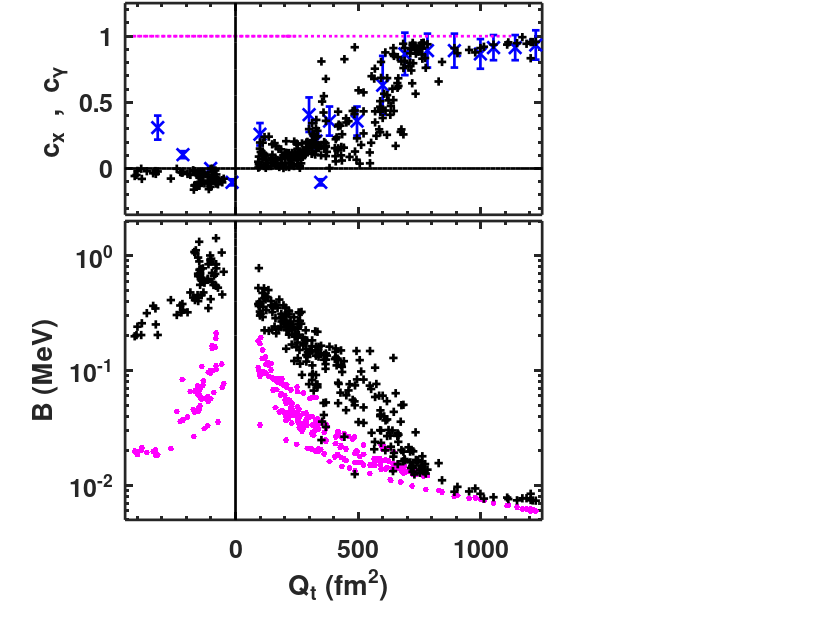} 
\caption {Correlation between the transitional quadrupole moment $Q_t$ on the x-axis and the axiality parameter $c_x$ as derived from data on $R_{42}$ (upper panel). A comparison to values derived from theoretical calculations on triaxiality \cite{de10} are shown in blue x with uncertainty bars for randomly selected $Q_t$ in the top panel as well. In the lower panel the excitation energy  parameters B are shown as derived from the measured $E_x(4^+)$ using the $c_x$ and Eq.~\ref{eq:1}. Experimental data are depicted as black crosses (+) and the magenta squares represent rigid rotation ($c_x$ =1); in the lower panel the correction for broken symmetry from Eq.~\ref{eq:2} was applied. }
\label{fig:1}
\end{figure}

We then accept the lowest observed $4^+$ states as good measure of the inertia term $B$. Using our axiality parameter~$c_x$ we get:
\begin{equation}
  B = \frac{E(4^+)}{4+16\cdot |c_x|} \simeq \frac{1.0 \cdot \hbar^2}  {|c_x| \cdot \Im_{rr}}; \   \
  \Im_{rr} = \frac{2}{5} A  m_N \cdot ( R_0^2 + 0.41 \frac{Q_t}{Z})\ ;  
\nonumber 
\end{equation}
\begin{equation}
 Q_t = sign(R_{42}-2) \cdot \sqrt {\frac {16 \cdot \pi}{5}
 B(E2,0 \rightarrow 2)} 
\label{eq:2} 
\end{equation}
Nuclear mass A and charge Z only appear via reference to the rotor moment of inertia $\Im_{rr}$ of the respective nucleus. The popular theoretical prediction \cite{al56} for it is inspired by macroscopic physics and based on axial symmetry.  In Eq.~\ref{eq:2} it is multiplied by $c_x$ to quasi define a parameter $\Im_{eff}$, which is smaller than the rigid rotor one. Here the nucleon mass is denoted by $m_N$ and $R_0 = 1.2\cdot A^{1/3}$ stands for the nuclear radius.
In Fig.\ref{fig:1} the non-axiality $c_x$ and the inertia parameter $B$ are depicted versus $Q_t$ which we use as x-axis. One sees a surprisingly good correlation between these two observables, both considered related to collective motion. And we find $c_x$~$\rightarrow$ 1 for large $Q_t$ and thus deformation. The alternative for rigid axial rotation is also depicted and it shows an irregular scatter for small $Q_t$. 

\  \

Our – heuristic – ansatz presented in Eqs.~\ref{eq:1}~and~\ref{eq:2} and depicted in Fig.\ref{fig:1} serves two aims: 

(1) It presents an empirical correlation between the observed B(E2,0 $\rightarrow 2)$, i.e. the $Q_t$, and our indicator of broken axiality -- derived from $2^+$ and $4^+$ level energies. The corresponding modification of the rigid rotor formula leads to a semi-empiric expression for $E_{yr}$. 

(2) It also can be regarded an analytic representation of results from the above-mentioned HFB/GCM calculations \cite{de10} regarding nuclei in the valley of stability. 

The use of $Q_t$ in Fig.\ref{fig:1} for both aspects is justified by the fact that the HFB/GCM-calculations agree well to experimental B(E2)-values. In the top part of it we  find reasonable agreement of $c_x$ derived from the experimental $R_{42}$ (see Eq.~\ref{eq:1}) to $c_\gamma = cos(3\gamma)$, tabulated as taken from CHFB/GCM \cite{de10}. In spite of the rather large uncertainties we consider the obvious agreement as a hint to assume them to be nearly equivalent. In the lower part of the figure the inertia parameters B is depicted in its dependence on $Q_t$; it was derived from the $4^+$ level energies using again Eq.~\ref{eq:1}. This makes its variation with deformation clearly stronger than for axial rotation with $\Im_{rr}$.

In both panels a clear correlation with $Q_t$ is indicated for the strongly deformed lanthanide and actinide  nuclei with $Q_t > 700~$fm$^2$ and for $Q_t \leq 300~$fm$^2$ there is a strong correlation as well. Also for intermediate deformation a trend of $c_x$ rising (and $B$ falling) with $Q_t$ is observed, although a significantly increased scatter is seen. Shell model configurations of smaller $c_x$ may lead 
to aligned bands with related states at lower energies than the 
rotational excitations considered here and and in a few cases this disturbs our simplified view. Similar arguments hold for a small portion of nuclei showing $R_{42}<2$ -- which are depicted for information as they suggest negative $Q_t$. We found out that the nuclei in both groups have preferentially $ 84 < A < 100 $ and $ 200 < A < 220 $ and  shell effects may become important there; this may find attention in future work.  

The quite smooth dependence of the two data sets $c_x$ and $B$ on $Q_t$ and the reasonable agreement to the prediction \cite{de10} suggests further theoretical work. Recently it was shown \cite{al17,sc19}, that for 12 nuclei with $R_{42}>2.7$ sufficient experimental data are available to perform a fully 3-dimensional analysis with 3 moments of inertia and 3 deformation parameters ($\gamma >0$). Our 1-dimensional approximation for more than 300 heavy nuclei will show that axial symmetry is broken quasi generally. An independent justification of this and of identifying $c_x$ and $ c_\gamma = \cos (3\gamma)$ was found by regarding old calculations \cite{df58, dr59} of rotational motion for nuclei which do not possess axial symmetry. The rather similar shape of the curve above $Q_t$ = 800~fm$^2$ to one derived from the conventional rigid rotor expression \cite{bo75, al56}, without any consideration of symmetry breaking, does not deliver a counter-argument: for strongly deformed nuclei axial symmetry was demonstrated \cite{mo59} to be nearly conserved. In the other extreme Eq.~\ref{eq:1} would result for $^{208}$Pb in $\gamma \simeq 33^\circ$ and hence of slightly oblate shape -- instead of spherical as assumed usually \cite{po20}.

 \
 
From the experimental $R_{42}$ we can also deduce the slope $E_x$ versus I (cf. Eq.~\ref{eq:1}) and in the following we will show that in a well deformed heavy nucleus our approach agrees rather well also to data for higher spin. 

\begin{figure}[ht]
\includegraphics[width=1.5\columnwidth]{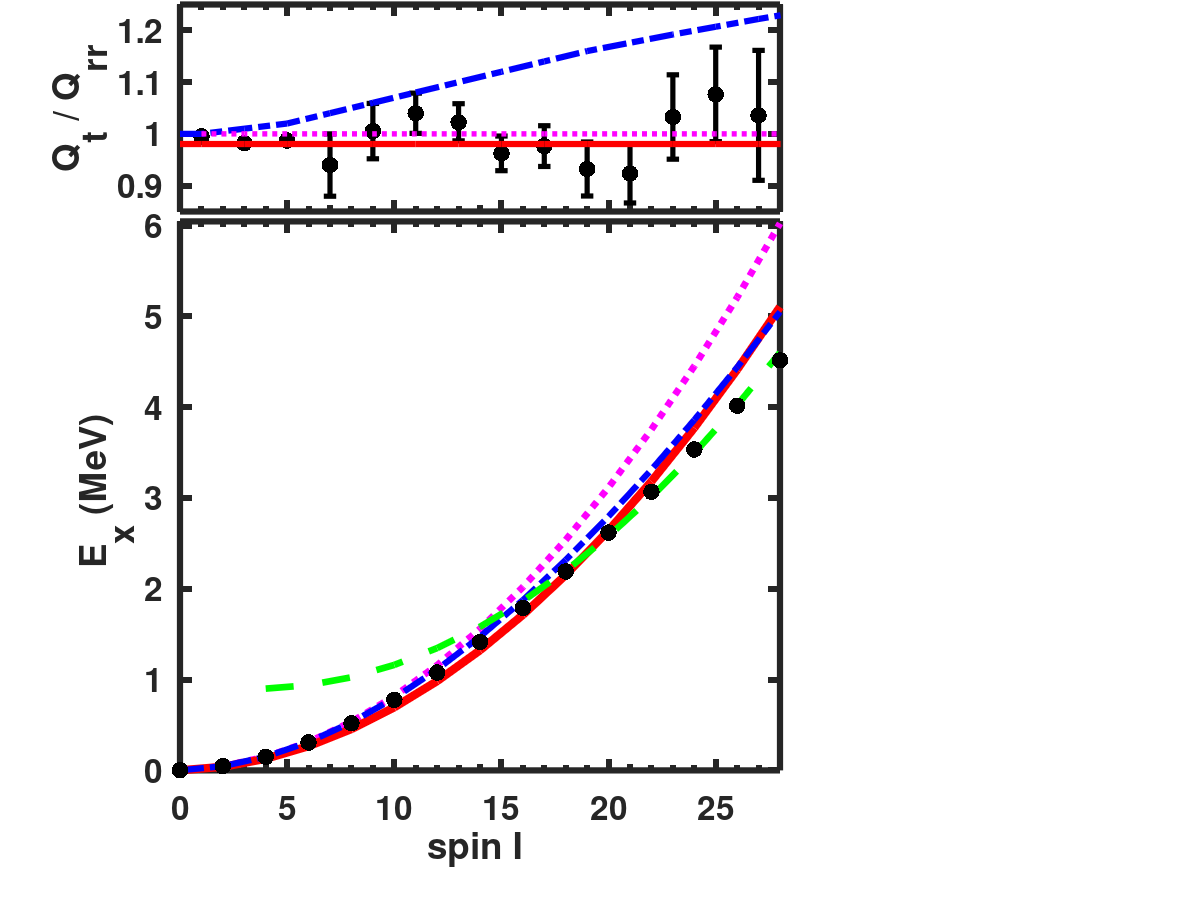} 
\caption{Experimental information \cite{gr81} is shown for the yrast band of $^{238}$U. In the upper panel B(E2)-values are depicted in their spin dependence and normalised to the rigid rotor value \cite{al56}. In the lower one the experimental excitation energies (black dots) are shown.  Data are compared to our proposal for a dependence on $c_\gamma$ (red curve), cf. Eq.~\ref{eq:1}, with $\gamma= 8^\circ$ taken from a HFB calculation \cite{de10}. Up to spin $I=18$ the ground band is well reproduced by a constant moment of inertia of $\Im = 64\hbar^2$ MeV; this differs from $\Im_{rr}$ in accord to Eq.~\ref{eq:2}. Depicted as well are the axial rigid rotor (magenta dots) and RVM calculations (dash-dot in blue) \cite{ob77}. Apparently another (green dashed) band starting near 1 MeV and angular momentum $I = 4$ is crossing near spin $I = 20$; for this band in Eq.~\ref{eq:1} the same $\Im$ and an increase to $\gamma=15^\circ$ were used, decreasing $c_x \simeq c_\gamma$=cos(3$\gamma$) by 23$\%$.}
\label{fig:2}
 \end{figure}

For $^{238}$U a multiple Coulomb excitation study was performed \cite{gr81} to learn about a possible rotational stretching. This was discussed \cite{ei87} and related to the  rotation-vibration coupling model (RVM) \cite{ma69}. At variance to these ideas no increase of the quadrupole strength within the yrast band is observed: The upper panel of Fig.~\ref{fig:2} depicts intrinsic quadrupole moments in the yrast band and the independence on $I$ is obvious up to $I^\pi = 28^+$. Their experimental determination has been described in detail \cite{gr81} and the negligibly small influence of axial symmetry breaking \cite{df58, dr59} on these was accepted. Apparently the dependence of $E_x(I)$ on triaxiality proposed in Eq.~\ref{eq:1} works also for higher spin, at least up to eventual band crossings. Our conclusions have been confirmed in experiments for neighbouring nuclides and similarities have been found in data for other heavy nuclei \cite{ow82, em84, ku89, pi93}. Also there the rigid rotor like dependence of $Q_t$ on spin I, i.e. the absence of rotational stretching was observed as well as the deviation of $E_{yr}$ from I(I+1). The observed irregularities could be explained as due to crossing bands, possibly of non-collective origin.

From Figs.\ref{fig:1} and \ref{fig:2} it is obvious, that our heuristic ansatz to account for the broken axiality in heavy nuclei allows to get a rather good description of the yrast band energies and of their dependence on the quadrupole deformation without the need for a local fitting per nuclide. In Eq.~\ref{eq:2} we use a global "effective moment of inertia" $\Im_{eff}=|c_x| \cdot \Im_{rr}$ to calculate the yrast energies and hence multiply the rigid rotor value and the non-axiality parameter $c_x$, which we derive from experimental $R_{42}$ values using Eq.~\ref{eq:1}.  Apparently our triaxial approach helps to understand the observed deviations from assuming a rotating axial nuclear body better than attempts to add vibrational degrees of freedom  -- although additional fit parameters are applied there. 

Only for well deformed nuclei with $Q_t > 750$ fm$^2$ (i.e. lanthanide and actinide nuclei) the $E_x(I)$ are rather close to the calculation for $c_\gamma = 1$ and this has often been considered a justification for treating such nuclei as axial. However, Fig.~\ref{fig:2} shows that even in $^{238}$U deviations from a non-axiality are seen in experimental data. For the lighter of the nearly 300 nuclei, for which $R_{42}$ is tabulated \cite{pr17}, this ratio is clearly below the "rotational" value 20/6=3.333. And our heuristic way to derive an axiality parameter $c_x$ from them allows to quantify the departure of yrast energies from the I(I+1) rule as well as it leads to a  change of the absolute value for moments of inertia. 

Here it is worth to recall that by an unconventional analysis of multiple Coulomb excitation data axiality breaking was detected quasi-experimentally \cite{wu96} for 25 well deformed nuclei; there electromagnetic transition rates to non-yrast states were included. But our approach indicates axial symmetry breaking to be a general feature of all heavy nuclei in the valley of stability. The apparently still controversial situation indicates the need to inspect other observables with possible sensitivity to broken symmetry -- as we will discuss now.

\section{Level and State Densities}
\label{sec:3}

If in a region around $E_x$ all levels are known, their average spacing $D$ determines the level density $\rho(E_x)=1/D$. From the level spins I the density of all magnetic sub-states is derived by $\omega(E_x)=(2I+1)\cdot\rho(E_x)$. Nuclear state densities are important inputs to calculations of compound nuclear reaction yields and for the analysis of gamma-decay cascades. For their determination one fact has to be considered: Like in other many-Fermion systems a transition is expected \cite{gr85,ig93,de03} from a Fermionic phase at higher energy to a Bosonic region dominated by pairing. The point between the two phases is defined by a transition temperature $T_{pt}$ which is well known from Fermi gas theory. In nuclei it is related to the gap parameter by $T_{pt}= 0.567 \cdot \Delta_o$ and the expression $\Delta_o=12/\sqrt{A}$;  another A-dependence exists in $\tilde{a} = \frac{\pi^2 A} {4 \epsilon_F}$ with $\epsilon_F=\sqrt{m^2_N+p^2_F}-m_N$ and here we use $p_F=$ 240 MeV.

\

In the following we use these relations together with modifications to the Fermi gas model (FGM) used frequently \cite{be36,er58,gi65,hu74,bj73,gr85,do19}. With these an approximation valid for a homogeneous Fermi gas can be derived to result in an analytic form to calculate nuclear state densities using Eq.~\ref{eq:3}. This equation results from Eq. 2-125 in ref. \cite{bo69} and an influence from pairing, shell structure and other effects is accounted for by the use of $M_{gs}$ in the determination of $E_{bs}$. 

From the different formulations for state density predictions important facts can be worked out which enter the modifications of the Fermi gas expressions as we use them: 

1. In a thermodynamical  approach the nuclear entropy S is related to energy U by $ S = 2 \sqrt{\tilde{a}\cdot U}$ via $\tilde{a}$, which hence is no longer considered only a level density parameter to be adjusted freely; 

2. U in the Fermi gas expression is back-shifted by $E_{bs}$ relative to the nuclear excitation energy $E_x$; 

3. collective enhancement and group theoretical factors account for the symmetry of nuclear shapes and their rotation in space; as outlined in Eq.~\ref{eq:1}, Eq.~\ref{eq:2} and Eq.~\ref{eq:4} the parameters  $c_x$ and $B$ enter here via $E_{yr}$.

Before we later treat the enhancement related to shape symmetry breaking -- the topic of this publication -- we start by discussing $\tilde{a}$ and $E_{bs}$ as appearing in the following expression for the density of states in the Fermi gas of neutrons and protons: 
\begin{equation} 
\omega_{np}(E_x) = \frac{\sqrt{\pi}\cdot e^S}
{12 \cdot\tilde{a}^{1/4}\cdot U^{5/4}}; \  \ \tilde{a}= \frac{\pi^2 A} {4 \epsilon_F} \simeq  \frac{A}{12.2 \ MeV} 
\nonumber \end{equation}  
\begin{equation}
E_{bs} = M_{mac} - M_{gs} ; \  \ U = E_x - E_{bs}; \  \  S = 2 \sqrt{\tilde{a}\cdot U}
\label {eq:3}  \end{equation} 

 In the spirit of our schematic approach we define the back-shift energy $E_{bs}$ by the difference between a macroscopic theoretical mass $M_{mac}$ and the measured ground state mass $M_{gs}$. Thus we do not need to use theoretical predictions for pairing, shell and other effects. And $M_{mac}$ is assumed to define the temperature zero of the gas phase and the Fermi gas energy $U$ is shifted accordingly, as shown in Eq.~\ref{eq:3}. Different from a similar  approach \cite{me94} and from our previous study \cite{gr14}, we adapt a Thomas-Fermi calculation (TF) \cite{my96} and set  $M_{mac}= M_{TF}$. We note that in our semi-empirical approach even-odd correction and the Wigner term are part of $M_{mac}$. At variance to LDM fits the TFC leads to positive $E_{bs}$ as it predicts $M_{mac}$ to be larger than $M_{gs}$ -- at least for nuclei not too far from the valley of stability.

\begin{figure}[ht] \includegraphics [width=1.03\columnwidth] {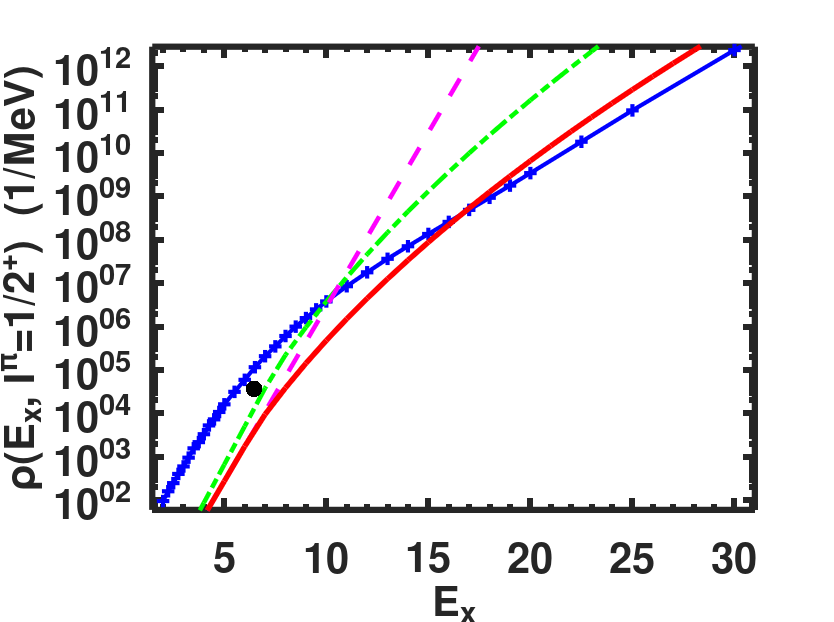}
\caption{Average level densities $\rho(E_x , I=1/2)$ in $^{161}$Dy in dependence of excitation energy. An experimental data point taken near $S_n$ is shown together with results of three calculations, all in absolute scale without any normalisation: Blue crosses on a line depict HFB plus combinatorial calculations \cite{gy08}, the drawn line in red results from our calculation approximating the Fermi gas model (FGM), for which we use $\tilde{a} = A/12.2$ as explained in the text. The dash dotted green line shows the effect when changing this small “level density parameter" $\tilde{a}$ to $A/10$ as proposed earlier \cite{ig93,ju97}. The magenta dashed line depicts an extension of the constant temperature part to $E_x$ above $E_{pt}$=7.4 MeV.}
\label{fig:3}
 \end{figure}

At and above $E_{pt} = \tilde{a} \cdot T_{pt}^2 + E_{bs}$ the state density is taken from Eq.~\ref{eq:3}  representing our approximation to the FGM . Hence it is dominated by the parameter $\tilde{a}$ and the back-shift energy $E_{bs}$ = $E_x-U$. In Eq.~\ref{eq:3} both appear in the exponent and have similarly large influence. We will first show that allowing axial symmetry to be broken $\tilde{a}$ can be set as small as $A/12.2$ MeV$^{-1}$, in contrast to previous work \cite{gi65, mu93, ig93}. This is demonstrated in Fig.~\ref{fig:3} by comparison to a recent combinatorial calculation \cite{gy08, ca09}. At high $E_x$ our approach compares to this calculation much better than the CTM-extrapolation. It also disagrees to the model with enlarged $\tilde{a}$, especially when the often used normalisation to the data point at $S_n$ is applied. As will be shown in the following, the resulting reduction in $\rho(S_n)$ is compensated by allowing axial symmetry to be broken.
 
 \ 
 
Presenting principal arguments \cite{gi65} it was proposed long ago to use Fermi gas expressions only above a critical energy and to complement it by an exponential interpolation below. Such exponential energy dependence -- often called constant temperature model (CTM) --  circumvents the difficulty of an extension of the FGM down to zero energy \cite{gr85}.  We also use this heuristic approach of a composite prescription \cite{gi65} but we identify the critical energy by $E_{pt}$ corresponding to the phase transition: $E_{pt}=\tilde{a} \cdot T_{pt}^2 + E_{bs}$. Semi-empirically we use in the ground state region an interval with the upper energy $E_x(4^+)$ and assume it to contain 3 levels. Thus we use $\omega(E_x) = 18/E_x(4^+) \cdot \exp(E_x/T_{eff})$ and adjust $T_{eff}$ to $\omega(E_{pt})$. We thus avoid a sudden change at $E_{pt}$ but we allow a change in the slope of $\omega(E_x)$, as thermodynamics predicts such a change in entropy at a $2^{nd}$ order phase transition. This means that $T_{eff}$ is not equal to $T_{pt}$.

We now treat symmetry effects neglected so far. To our knowledge a Fermi gas ansatz without shape symmetry was not yet compared to experimental nuclear level densities and we will present a scheme to perform such a test for 150 heavy nuclei.
As will be shown as well, axial symmetry breaking becomes important with the transformation to level density $\rho(E_x)$ by Eq.~\ref{eq:4}.
Uncorrelated nucleons in the Fermionic gas form states with angular momentum K. These may couple to levels of correlated collective bands -- mainly rotations of the entire nucleus within its environment. Around $E_x = E_{pt}$ the level density is enhanced by coupling all such bands to every intrinsic state. This was analyzed for a non-spherical body with axial symmetry \cite{er58, hu74, do74, do19}. But, as was predicted \cite{bj73, ha83}, accepting its breaking results in a larger rise. Thus collective enhancement arises in the transformation of the intrinsic state density $\omega_{qp}(E_x)$ to the observable level density $\rho(E_x)$. To quantify this we follow a suggestion made by Bohr and Mottelson in their textbook on deformed nuclei \cite{bo75} (cf. Eqs. 4-63 to 4-65 and the related text). They argue that by rotation of an intrinsic body without axial symmetry every state with K becomes a member level (with spin I=K) in one of (2I+1) bands \cite{bj73, do19}. The resulting enhancement is given by the factor $\sum_{I}(2I+1)$. But the rotational energy has to be subtracted; this is realised in a local temperature approximation by a multiplication of the intrinsic state density $\omega_{qp}(E_x)$ with an exponential cut off \cite{bj73,ha83}. We base it on the ratio $E_{yr}(I)/T_{pt}$ (with $T_{pt}$ as defined already) to obtain an approximation for the spin dependent level density at $E_{pt}$ and above:  
\begin{equation}
\rho(E_x,\pi=+) \simeq\sum_{I}\frac{(2I+1)}{4 } \cdot exp(-\frac{E_{yr}(I)}{T_{pt}})\cdot\omega_{np}(E_x); \ \ \
\label{eq:4} \end{equation}

The conservation of $\mathcal{R}$-symmetry and parity was included in it by a factor 1/4. But, if both are broken simultaneously in octupolar shapes \cite{gr75}, it rises to 1/2 and this may enlarge $\rho$ for certain A. The broken axiality is already accounted for by the derivation of Eq.~\ref{eq:4} and for the calculation of the exponential spin cut off we use Eq.~\ref{eq:1} and Eq.~\ref{eq:2}. Hence large $\Im_{eff}$ and small $\frac{E_{yr}(I)}{T_{pt}}$ result in a rotational enhancement and a peaking of $\rho(E_x,I)$ near to angular momentum $I \simeq 4$, which is the level spin we used to extract $B$ and $\Im_{eff}$ from observation. Our approach allows an application also for nuclei with small $Q_t$ and $\Im_{eff}$ for which the spin cut off becomes less important as they have larger $E_{yr}$. It is stressed here, that shape deformation numbers only appear in this cut off term indirectly via the correlation depicted in Fig.~\ref{fig:1}. Our scheme picks the inertia from the $0^+, 2^+ $and $ 4^+$ energies near the ground state and forms a logarithmic interpolation from the state density there to the Fermi gas value at $E_{pt}$. Neglect of additional modes like the one seen in Fig.~\ref{fig:2} makes this an approximation. 

\begin{figure}[ht]
\includegraphics[width=1.08\columnwidth] {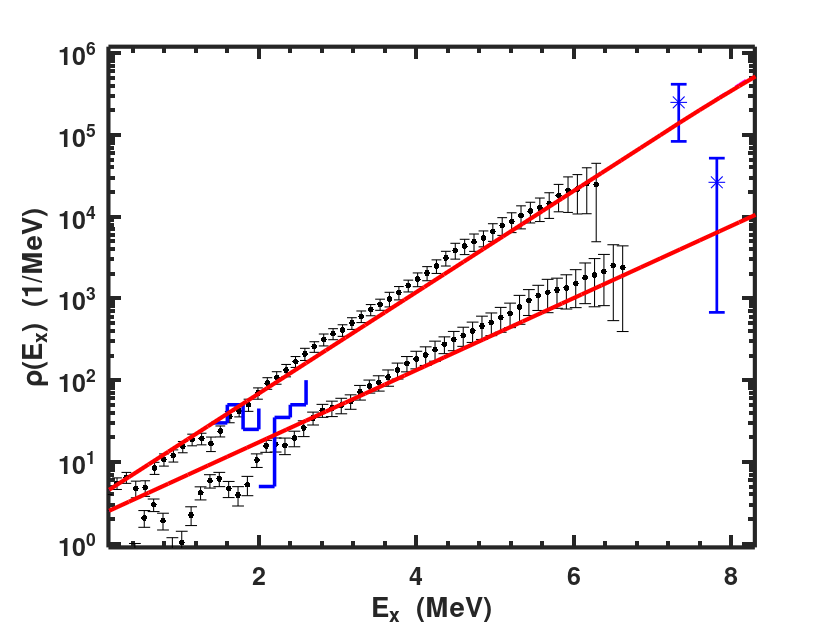}
\caption{ 
Total level density $\rho(E_x)$ in $^{144}$Nd and $^{148}$Nd (top); our predictions depicted by the red line are compared to experimental data from low energy levels and n-capture resonance spacings \cite{ca09,ig09} (marked in blue). The black data points stem from proton-photon coincidence yields observed at the UiO-cyclotron \cite{gu21}; they were normalized to the other data.} 
\label{fig:4}
 \end{figure}
For a comparison to experimental data in dependence of $E_x$ we use $^{144}$Nd and $^{148}$Nd for which a recent theoretical study \cite{gi18} predicted broken axiality. For these isotopes experimental level density information was derived at the Oslo cyclotron \cite{gu21} from coincident yields of photons cascading down from excited states of known energy. A regard of such data from photon yields justifies our use of the CTM below $E_{pt}$ \cite{gu15}. But a comparison on absolute scale is irrelevant as these data are adjusted to other observations which were analysed under the assumption of axiality; so we decided to repeat it. Oscillatory structure in these observations may indicate an underlying shell structure at small excitation which is not included in our scheme based on the assumption of equidistant elementary levels. 

We find a reasonable prediction for the absolute value of the level density in the region of  $S_n$ continuing down to 2 MeV  not using any parameter fit to the data. With our scheme we can identify the collective enhancement being reduced due to the  exponential cut off in Eq.~\ref{eq:4}. The reduction near magic shells results from a smaller backshift in Eq.~\ref{eq:3}. Already in the two isotopes $^{148}$Nd and $^{144}$Nd this is seen and the good agreement to data is considered a proof of our scheme based on broken axiality, on a phase transition at $T_{pt}$ and on the TF mass prediction \cite{my96}; it avoids any adjustment to local data other than $E_x(4^+)$ and $E_x(2^+)$.

\begin{figure}[ht] 
\includegraphics[width=1.08\columnwidth]{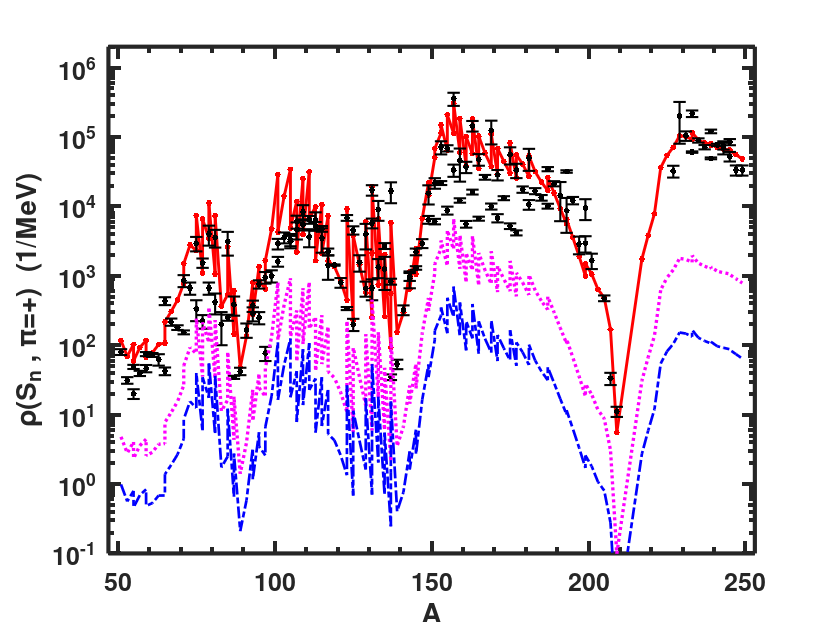}
\caption{Observed average level densities $\rho(S_n ,1/2^+)$ in nuclei with $51<A<253$. Compiled data (black dots) \cite{ig09, ca09} are compared to the prediction for $\rho(S_n,\pi = +)$ as discussed in the text: The curves depict the increase of $\rho$ in going from a situation without rotational modes (lowest dashed line in blue), the case of assuming axiality (medium dotted line in magenta) to the enhancement caused by giving up spherical and axial symmetry (highest line in red).}
\label{fig:5}
 \end{figure}

To give a broader overview on these effects we present in Fig.~\ref{fig:5} our prediction together with observations for levels in 150 odd nuclei with $I^\pi = 1/2^+$. These data near $S_n$ were obtained by the use of resonance spacings from neutron capture by even target nuclei. This choice delivers a large data base with the advantage of the same spin in all nuclei. Unfortunately for many of them no information is available for the angular momenta near ground and we have to refer here to the even neighbour nuclei. But the choice made for small $E_x$ does nearly not effect $\rho(S_n)$, as $S_n$ is nearer to the phase transition at $E_{pt}$. As is obvious in Fig.~\ref{fig:5} the experimental level density data show a continuous increase with $A$ and a significant local decrease near magic numbers. The experimental data for deformed nuclei are enlarged and the very strong increase from the lead region to the actinides \cite{pr10} is a challenge to any model. 
In our new approach, which is modified from work presented earlier \cite{be11, gr16, gr17, gr19}, the changes result from deleting an extra shell correction in ã as used in Eq.2 of \cite{gr17}, a decrease in the spin cut-off effect and also the newly defined back-shift energy $E_{bs}$ which is of influence on $E_{pt}$. In the regions of magic numbers its rise is due to pairing and shell effects lowering the ground state mass and that also induces an increase of $T_{eff}$ in the CTM part. When preparing Fig.\ref{fig:5} we realised that low spin states as reached in neutron s-capture by spin 0 target nuclei are the main component in Eq.~\ref{eq:4}.

As depicted, a breaking of axial symmetry leads to a reasonable agreement for heavy nuclei, for which experimental data exist. Without fitting $\tilde{a}$ we arrive at a good accord in the valley of stability up to A = 250 -- on absolute scale over many orders of magnitude.  The small under-prediction of $\rho(S_n)$ for $A>200$ outside of $^{208}$Pb may indicate degrees of freedom not yet taken into account by our approach. 
In Fig.~\ref{fig:5} we also depict level densities obtained for unbroken symmetries as reviewed some time ago \cite{bj73, do19}. We insert $\omega_{np}(E_x)$ from Eq.~\ref{eq:3} into their equations including our account for ground state shell effects by a backshift in energy. For the spin-dispersion parameter $\sigma_K^2 = 1/\pi \sqrt{6\cdot\tilde{a}<K^2>T_{pt}}$ we follow the proposal \cite{gi65} to use $<K^2>$ from Fig.2 of an old Thomas-Fermi calculation \cite{je52}. The spin  dispersion depicted there depends on mass and charge number only and under the assumption of spherical symmetry it reduces $\rho(E_x)$ through division by $\sqrt{8\pi}\cdot\sigma_K^2$. In a recent study \cite{do19} an extra factor $\sigma_K$ was shown to be compensated by the integration over I. The spherical prediction for level density is far below experimental data as is known since long \cite{be36} and mainly due to the spin dispersion factor $\sigma_K^2$ appearing  in the denominator \cite{bj73, do19}. We note  that the apparent sphericity of ‘magic’ nuclei is not an exact symmetry; it results from a very small quadrupole deformation with the three axes nearly identical \cite{po20}.

Under the assumption of axial symmetry the FGM prediction is lower by $\sigma_K$ and this is  depicted in Fig.~\ref{fig:5} as well. The study mentioned already \cite{do19} uses a different approximation for $\sigma_K$, which is smaller by up to 10 $\%$, and it neglects back-shift effects. These are rather small in the 7 odd well deformed nuclei for which it presents a comparison to resonance spacing data. Hence the resulting average under-prediction by 3.8 does not deviate from our axial result for these nuclei. But, at variance to that work, we can  present results in a wide range of heavy nuclei and equally cover 'spherical', 'intermediate' as well as the 'deformed' regions. In all these a limitation to axiality leads to a reduction by more than an order of magnitude (for A$ >$ 150). This results mainly from the factor $\sigma_K^{-1}$ caused by the quest for axial symmetry of the intrinsic shape which does not appear under the assumption of broken symmetry. We are confident that the ongoing discussion about rigid or dynamic triaxiality is of no impact on our scheme. Independent of a deviation from zero for a triaxiality parameter its distribution with nonzero variance is equal to a new degree of freedom causing additional excited bands and thus higher level density.

As mentioned already, we use the low value $\tilde{a} \simeq \frac{A}{12.2}$ at variance to previous work \cite{bo69,hu74,ko08}. There $\tilde{a}$ was used as fit parameter for the level density observations and preference was given to $\tilde{a} \simeq \frac{A}{9}$ or even larger; this choice is in contrast to the value used in nuclear matter studies as selected to have stable very heavy nuclei and even infinite matter (Eq.2-125a in \cite{bo69}). We point out that the decrease resulting from the low value for $\tilde{a}$ is mainly compensated by allowing a sum in Eq.~\ref{eq:4}, giving up the request of an intrinsic axis for all the nuclei regarded, independent of their $Q_t$. This also results in the suppression of the axial I(I+1) rule in the exponential. Together with the use of $T_{pt}= 0.567 \cdot \Delta_o$ \cite{gr85} as fix-point, we obtain a surprisingly good agreement of our level density scheme to observation with no fit of parameters. Those of importance like $R_{42}, E_x(4^+),  \Delta_o, p_F,  E_{bs}$ can be derived from other experimental information and we arrive at a global prediction without any other local adjustment. Later we will apply this nearly parameter free prediction for $\rho(E_x)$ to the study of radiative strength.

\section{Giant Resonances and Broken Axiality}

\label{sec:4}
Previously, we have demonstrated how the breaking of axial symmetry becomes obvious from an analysis of experimental shapes of isovector giant dipole resonances (IVGDR) and we obtained good fits \cite{er10, ju08, ju11} for a number of isotopes. There we went beyond the fact, discussed long ago, \cite{ca71, be75} of effects induced by a breaking of sphericity. We regarded many nuclei with our triple Lorentzian (TLO) approach \cite{be11} using three poles with resonance energies reflecting the three nuclear axes, avoiding ad-hoc assumptions on axiality. We extended \cite{gr14, gr16} this concept making use of calculations \cite{de10} predicting deformation parameters which indicate the breaking of axiality for nearly all heavy nuclei in the valley of stability. We follow here earlier work for triaxial configurations as applied to giant resonance shapes and widths \cite{bu91}; these propose to use the Hill-Wheeler expression to relate the three pole energies $E_r$ to the mean resonance energy $E_o$. This quantity is taken from an expression derived long ago \cite{my77} including the standard parameters for symmetry energy $J=32.7$ MeV and surface stiffness $Q=29.2$ MeV \cite{mo08}. The apparent width as seen in the cross section is governed by the split induced by deformation; the axial symmetry breaking fixes the position of the middle pole. The spreading of the three resonances is shown to follow a power law, which we now write as $\Gamma_{GDR} \simeq  (E_r/6.95)^{1.6}$ MeV; this makes the base dimensionless. As shown there the value 1.6 follows from the wall formula \cite{bu91} and we have shown in previous work \cite{be11, gr16} that the value 6.95 MeV results in a good fit for nearly all heavy nuclei for which good data are available. It corresponds to a smaller spreading as predicted earlier and in contrast to theory \cite{my77, bu91} it does not vary with A. Together with the nucleon effective mass $m_{eff}$ set to $m_{eff}=800$ MeV we have two global parameters to predict the IVGDR cross section for $78<A<239$. Our concept could be shown to be global \cite{ju11, gr16}; to improve the fits for some heavy nuclei one may reduce their predicted quadrupole moment by some extent \cite{be07}.

Before a closer discussion of this point we mention to have used in the past a theoretical prediction \cite{de10} which is a generator coordinate mapping of a Hartree-Fock-Bogoliubov (HFB-GCM)-calculation with the Gogny D1S interaction. It should be mentioned here that its predictions agree reasonably well to observed $R_{42}$-values \cite{pr17} and, as already mentioned, it also predicts very well B(E2)-values for more than 300 heavy nuclei. To study an alternative we now discuss more sophisticated auxiliary-field quantum Monte Carlo (AFMC) simulations, which very recently also predicted axial symmetry breaking -- as published for a number of nuclei \cite{gi18}. At variance to the HFB/GCM calculations \cite{de10} they result in smaller quadrupole deformation, but larger variance. They find similar values for the breaking of axiality and we studied the effect of applying these deformation parameters in our concept of triple Lorentzians (TLO).

\begin{figure}[ht]
\includegraphics[width=1.5\columnwidth]{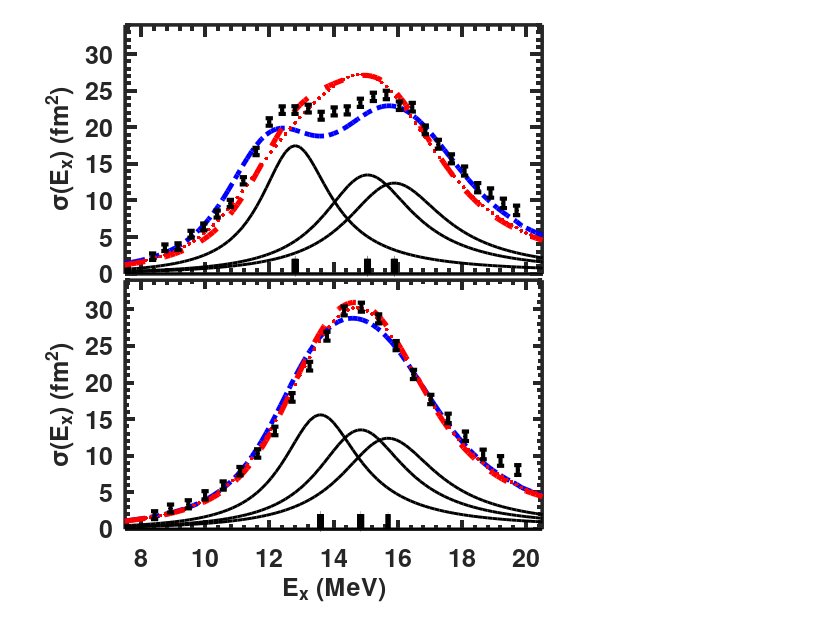}
\caption{ 
Cross section of photo-neutron production  on $^{152}$Sm (top) and $^{148}$Sm (bottom) \cite{ca74, di88} in comparison to the calculated sum for three Lorentzians (TLO) for the IVGDR. As discussed in the text global fit parameters are used together with deformation parameters from AFMC (red dashed line); the shapes and centroids of the three components are indicated in black and correspond to $Q_0\simeq 250 fm^2$ and $ 156 fm^2$, respectively. The dotted curve in red indicates the effect of the large fluctuation in the axial deformation as predicted (AFMC) for $\beta$; an approximation involving instantaneous shape sampling (ISS) \cite{er10} was used. The blue curve in short and long dashed line depicts the summed curve for the HFB-GCM calculation; it corresponds to $Q_0 = 371 fm^2 $. A beam energy spread of 0.7 MeV was assumed; yield outside of the IVGDR near 12 and 22 MeV had been identified \cite{gr16} to be part of giant quadrupole strength}
\label{fig:6}
 \end{figure}
In Fig.~\ref{fig:6} the results are depicted for two isotopes of Sm, selected because of their largely different experimental B(E2) and calculated deformation. For both the broken axiality \cite{bu91} results in a middle position of one pole and this makes the triple sum resemble a single resonance -- at variance to well deformed nuclei with $Q_t > 700 fm^2$, where the two upper poles are close. Here the axiality parameters $c_x = cos(3\gamma)$ derived from the two microscopic predictions are very close to each other for both nuclei; and they are not far from an evaluation using $R_{42}$. The good agreement of the resulting shape to experiment is considered an important proof of broken axiality, as it is achieved with a global (A-independent) width parameter, which fits all heavy nuclei with A $>$ 70 without any adjustment of a spreading width. But there is a fact to be pointed out: As the two microscopic calculations disagree in their predicted quadrupole deformation $\beta$ especially the for $^{152}$Sm a comparison to the observed IVGDR shape allows an evaluation of their closeness to experiment: A better agreement to data is expected from the use of an intermediate value for the quadrupole moment and it is very interesting to see that for $^{152}$Sm a hyperfine-structure study with muonic X-rays exists \cite{st05} which results in $Q_0(2^+) = 300 fm^2$ -- intermediate between the two microscopic calculations. As these do not publish them explicitly, the quadrupole moments had to be derived from the published deformation values using a cranking approximation \cite{ri82}; apparently there is need for more theory work, also for more cases.  

We stress a reasonable accord to the dipole sum rule and the overall shape as depicted in Fig.~\ref{fig:6}. The low energy slopes will in the next paragraph be shown to be of importance for the derivation of photon strength below $S_n$.

\section{Dipole Strength and Radiative Width}
\label{sec:5}

Radiative widths are important ingredients for the interpretation of nuclear data with photons in one of the channels. There have been many attempts \cite{mu18} in finding phenomenological expressions for photon strengths and widths to extrapolate from observations made in the valley of stability to neutron capture by exotic nuclei, like those involved in astrophysical processes. It was stated some time ago \cite{da64, bu91}, that wide doorway states like the IVGDR have a spreading width depending on the pole energy of the resonance and on the coupling to the underlying narrow states. 
For heavy nuclei it was proposed \cite{ba73} to extract E1 strength and energy dependence from the study of isovector giant dipole resonances (IVGDR). And in the following we will show that it is important to not assume either spherical or axial symmetry; the effect of giving up both was demonstrated in the last section with its Fig.~\ref{fig:6}. We have already demonstrated \cite{ju11, be11} the effect of broken axiality on the analysis of IVGDR data in very many heavy nuclei. It was shown how the extraction of strength from experimental data is influenced by a decomposition of the broad observed IVGDR distribution into the components related to nuclear deformation and strength spreading. The width finally deduced is the important quantity for a prediction of the low and high energy tails. As pointed out before \cite{gr16} TLO invoking three resonance poles leads to a reduced spreading width and this results in a low cross section outside of the main IVGDR peak; the total integrals are in a surprisingly good accord to the TRK sum rule \cite{ju11, be11}.

We will demonstrate a clear contrast of results from our TLO based analysis to recently presented single Lorentzian (SLO) fits \cite{pl11}. These, presented for many nuclei, disregard the different sources of the apparent IVGDR width and try to fit it by one or at most two Lorentzians. This point is also disregarded in the modification SMLO \cite{ko90} which was proposed in a publication \cite{ka83} with its $2^{nd}$ part often referred to as "KMF-model". To seemingly better describe radiative neutron capture data corresponding to lower photon energy it introduces the concept of dipole photon width to vary with the photon energy. We consider this part questionable as GDR's differ from narrow neutron capture resonances near $S_n$ (described in the $1^{st}$ part of that paper), where the widths of these vary with their pole energy $E_r$. And the related data analysis needs three independent local parameters for each nuclide. In addition, SLO and SMLO with their local adjustment of parameters produce a clear overshoot over this rule for many nuclei \cite{pl18}.  No explanation for this deviation is given although it indicates a clear disagreement to fundamental considerations \cite{ge54}. 

To relate the photon absorption cross section $\sigma_\lambda (E_x)$ and average gamma-decay width $\langle\Gamma_\gamma\rangle$ of radiative capture resonances at $E_x$ the concept of photon strength function $f_\lambda$ was introduced \cite{ba73, mu00}. A brief expression reduced to those  electric dipole excitations which are assumed to be enclosed in the IVGDR tails is given in Eq.~\ref{eq:5}. The left part describes the extraction of the strength function from the corresponding absorption cross section $\sigma_{E1}(E_x)$ of photons with energy $E_x$; the quantum-mechanical weight factor $g_{eff}$ was discussed earlier \cite{gr16}. The right part of Eq.~\ref{eq:5} regards dipole radiation to final states $f$. For the case of unknown energy the denominator contains a statistical average over all $E_\gamma$ possibly reaching all $E_f$, which we assume to cover the range from 0 to $S_n$. With detailed balance and the Axel-Brink hypothesis the photon absorption cross section is set equal to the average photon width normalized to photon energy $E_{\gamma}=E_x-E_f $ and the density of narrow resonances at $E_x$. Hence it is assumed that $f_1(E_\gamma)$ is the same for $\gamma$-absorption as well as for $\gamma$-decay and we approximate:
\begin{equation}
\dfrac{\sigma_1(E_x)} {(\pi\hbar c)^2 \ E_x \ g_{eff}} \simeq f_1(E_x)
\simeq \dfrac{\langle \Gamma_\gamma(E_x) \rangle\cdot\rho(E_x)} 
{\langle \omega(E_f) \cdot(E_\gamma)^3\rangle\cdot S_n}  
\label{eq:5}   \end{equation}
In the right part the average over $E_x$ in the numerator is taken from observations \cite{ig09, mu18} and this reduces fluctuation effects. In the denominator the expression $E_\gamma^3$ for the decay refers to the first photon in a cascade. Very often it is not measured and it is approximated here by a statistical average including $\rho(E_f)$; this leads to an increase as compared to just using $E_\gamma^3$. All level densities are calculated by allowing non-axiality as described above. 
\begin{figure}[ht]
\includegraphics[width=1.05\columnwidth]{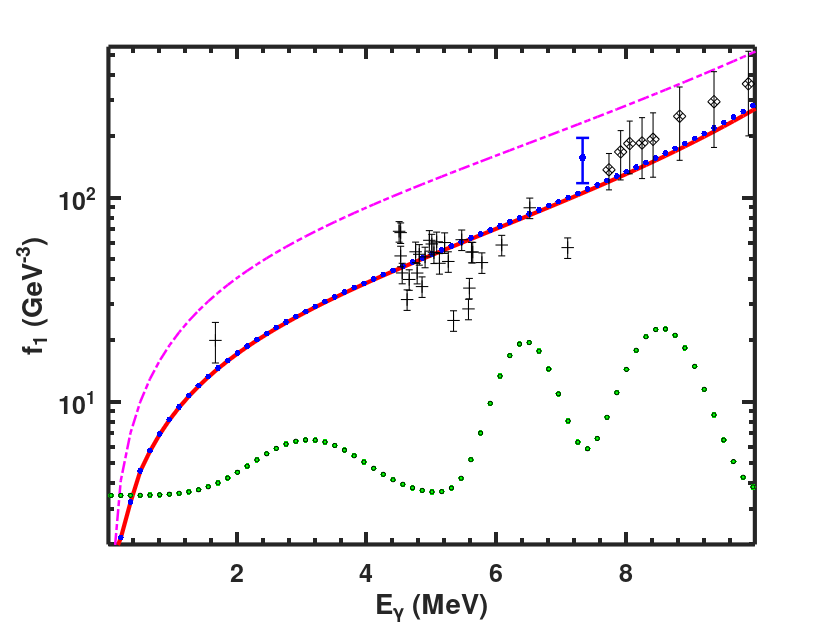}
\caption{ 
Dipole strength functions $f_1 $ are shown for $^{146}$Nd in their dependence on $E_\gamma$. For comparison predictions \cite{he10, gr16} for M1 are depicted by (o). For E1 good agreement appears between the dots and the drawn line as derived from the two types of calculations on display in Fig.~\ref{fig:6}, based on HFB-GCM \cite{de10} and (AFMC) \cite{gi18}. In contrast to a SMLO-prediction (-. in marron) there is good agreement to very new $(\gamma,n)$ measurements \cite{ny15} (black diamond) and to ANC-data described in the text (black +). An experimental value \cite{ig09} converted using Eq.~\ref{eq:5} (blue I) agrees also as well as an old estimate \cite{po75} near 2 MeV}
\label{fig:7}
\end{figure}
Fig.~\ref{fig:7} shows the case of $^{146}$Nd which allows a comparison to previous work \cite{mu00}, which had carefully studied details in the gamma spectra resulting from average neutron capture (ANC). They agree well with $f_{E1}$ resulting from the prediction obtained with global parameters as demonstrated in Fig.~\ref{fig:6}. As obvious, there is no need of any additional energy dependence or height adjustment to observation. We consider this a clear success of the TLO description of IVGDR strength already demonstrated in our global approach \cite{gr16}. There good agreement was shown also to the measured low energy dipole strength from recent photon emission and absorption data -- as long as such information was available. It has been applied to many nuclei with axial symmetry breaking as quantified by the calculations available at the time \cite{de10}. We propose to concentrate on TLO to derive $f_{E1}$ and to disregard other possible components.

The figure indicates that near the threshold for n-emission collective modes like magnetic spin flip, scissors etc. contribute little and thus are of minor influence on predictions for radiative capture. But attention is needed for low energy transitions $\leq$ 1 MeV from the capture region (c) to the quasi-continuum (q) part of the excitation spectrum. Such cq-transitions were observed long ago in $(n,\gamma \alpha)$ reactions and assumed to be of M1 type \cite{fu73, po75}. In Fig.~\ref{fig:7} we extended the  M1 component respectively as approximated by Eq. 2.57 of a recent review \cite{mu18}. The qq-transitions are followed by photons of $E_\gamma \leq S_n$ which have been observed in $^{146}$Nd \cite{ve99} and many other heavy nuclei \cite{be92, kr19}. They may erroneously be identified as originating from other modes eventually labelled as "pygmy" \cite{ba73} and we recommend more detailed studies -- theoretical as well as experimental.

\begin{figure}[ht] 
\includegraphics[width=1.05\columnwidth]{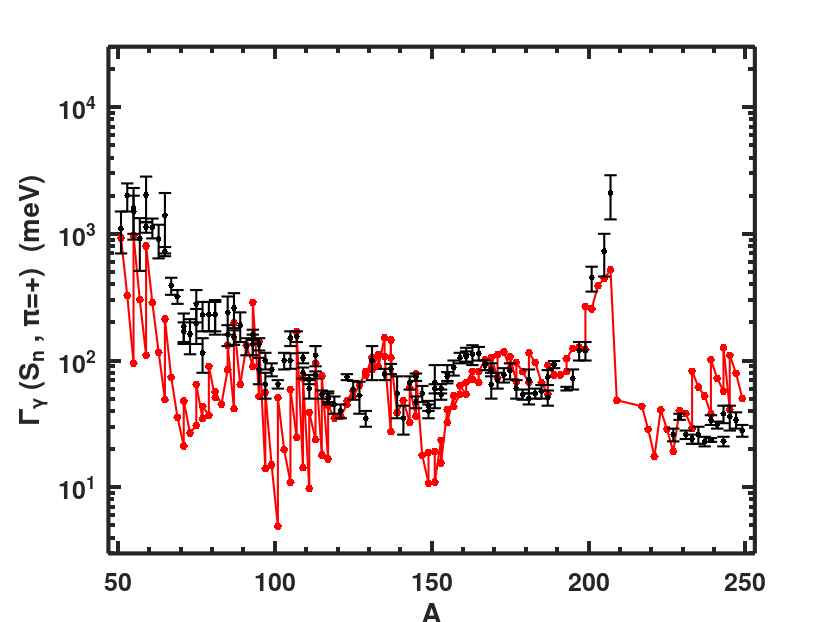}
\caption{ Average photon widths $\langle \Gamma_\gamma \rangle(S_n ,1/2^+)$ in nuclei with $51<A<253$ as observed in neutron capture by spin 0 target nuclei. Data (black dots) compiled \cite{ig09, ca09} are compared to the prediction as discussed in the text: The curves depict the variation of $\langle \Gamma_\gamma \rangle$ due to changes in state density induced by giving up axial symmetry and by shell effects (line in red).}
\label{fig:8}
 \end{figure}

To study the A-dependence of our scheme we have tried to extend our predictions to the well documented Maxwellian averages for n-capture \cite{gr14} but then we were not sure about the n-strength functions. This is why we now present in Fig.~\ref{fig:8} gamma-decay widths predicted on the basis of Eq.~\ref{eq:5}. There we used calculations of GDR cross sections based on our TLO-concept combined to our level density scheme presented in section 3. Reasonable agreement to observations is obvious from the figure: The general trend is reproduced as well as the sudden increase caused by the increase of backshift near closed shells.

It is stressed here that our predictions are on absolute scale and obtained without much fitting to observables and with very few free parameters. These are $m_{eff}$ and in the case of dipole strength only the TRK sum rule. In addition one globally valid spreading width parameter combined to the exponent 1.6, which was derived within the one-body dissipation model \cite{bu91}. The level density and photon width predictions have no other free parameters as $\tilde{a}$ is taken to agree to nuclear matter calculations and the phase transition point is derived from general Fermion pairing theory \cite{gr85, de03}. In both fields our work is completely based on broken axial symmetry. As mentioned above the triaxiality parameter $\gamma$ can be either found in microscopic calculations or through a heuristic interpretation of $R_{42}$-values as observed in quasi all heavy nuclei. 

\section{Conclusions}
\label{sec:6}
Broken axial symmetry was demonstrated to have consequences on various observables from different fields of nuclear physics. Axial symmetry breaking was first detected in experimental spectroscopy data for heavy odd nuclei. In the study of some even nuclei by multiple Coulomb excitation \cite{cl86} non-axiality was shown to result from an analysis in a rotation invariant way. Even for well deformed nuclei axiality was shown to be broken \cite{wu96}; there the variance of the triaxiality parameter $\gamma$ is centered is in the range of $\gamma \sim 8^\circ$. In accordance to recent HFB/GCM calculations\cite{de10} the value 0 is outside of the variance and we have shown: 
   
1. Allowing for broken axial symmetry by introduction of a newly proposed modification of the commonly used $I(I+1)$ rule higher spin yrast energies in the actinide nuclei are well reproduced up to the band crossing region. Here the excitation energies of the $2^+$ and $4^+$ is the only information used and we do not invoke an extra variation of the moment of inertia \cite{gr81}.
   
2. Due to reduced symmetry level density predictions are collectively enhanced as compared to assuming a symmetric shape. Assuming a break of axiality enhances them so much that the use of $\tilde{a}\leq A/12$ in the Fermi gas formula can be used -- a value agreeing to nuclear matter studies. We show it to strongly improve the agreement to level densities at excitation energies near $S_n$ as obtained from neutron capture resonance spacings.
   
3. For nuclei with intermediate deformation TLO allows a good representation of dipole strength data in the IVGDR range. As demonstrated, the false assumption of axiality requires three or more local parameters for every nucleus to obtain a reasonable fit \cite{pl18}. In contrast TLO allows Lorentzian fits in quasi all heavy nuclei with one global parameter for the spreading width. The resulting rather small width values allow agreement to the TRK sum rule and reduce the prediction of electric dipole strength at low energies.
   
4. Assuming broken axial symmetry in the equations for level density and dipole strength leads to a novel analytic and parameter free approximation of average widths for radiative neutron capture and we see a good agreement in the comparison to compiled observations.  

\

We consider our findings to be a falsification of the often made ‘ad hoc’ assumption of all heavy nuclei’s axial symmetry. They also make us warn to use the popular $I(I+1)$ rule for the spin dependence of ground band energies. Similarly we propose not to insist on the very often made distinction between spherical and deformed nuclei. We present a continuous transition reaching from small to large ground state quadrupole moment and we are aware that broken symmetry may complicate theoretical calculations. But we have shown that various observations clearly point in the direction of a breaking of axial symmetry. And we think that this is an experimental verification of the Jahn-Teller effect in heavy nuclei.  

\section*{Acknowledgements}
This work has been supported by the German federal ministry for education and research BMBF (02NUK13A), and by the European Commission through  Fission-2013-CHANDA (project no. 605203) and Horizon 2020 (ARIEL project no. 847594). Intense discussions within these projects and with other colleagues, especially with Roberto Capote, Hans Feldmeier, Stephan Goriely, Sophie Péru, Ronald Schwengner, Julian Srebnry, Jonathan Wilson and Hermann Wolter are gratefully acknowledged.


\newpage{\pagestyle{empty}\cleardoublepage}
%
%
\end{document}